\documentclass[preprint,showpacs,preprintnumbers,amsmath,amssymb]{revtex4-1}
\usepackage[colorlinks,linkcolor=blue,anchorcolor=blue,citecolor=blue]{hyperref}
\usepackage{graphicx,epsfig}
\usepackage{dcolumn}
\usepackage{bm}
\usepackage{fixfoot}
\usepackage{flafter}
\usepackage{etoolbox}
\usepackage{kbordermatrix}
\begin{document}
\title{ Exact solutions and symmetry analysis for the limiting probability distribution of quantum walks}
\author{Xin-Ping Xu$^{1}$}
\author{Yusuke Ide$^{2}$}

\affiliation{%
$^1$School of Physical Science and Technology, Soochow University, Suzhou 215006, China \\
$^2$ Department of Information Systems Creation, Faculty of Engineering, Kanagawa University, Yokohama, Kanagawa, 221-8686, Japan
}%
\begin{abstract}
In the literature, there are numerous studies of one-dimensional discrete-time quantum walks (DTQWs) using a moving shift operator. However, there is no exact solution for the limiting probability distributions of DTQWs on cycles using a general coin or swapping shift operator. In this paper, we derive exact solutions for the limiting probability distribution of quantum walks using a general coin and swapping shift operator on cycles for the first time. Based on the exact solutions, we show how to generate symmetric quantum walks and determine the condition under which a symmetric quantum walks appears. Our results suggest that choosing various coin and initial state parameters can achieve a symmetric quantum walk. By defining a quantity to measure the variation of symmetry, deviation and mixing time of symmetric quantum walks are also investigated.
\end{abstract}
 \maketitle
\section{Introduction}
Quantum walks (QWs) are analogues of classical random walks, designed primarily with the aim of finding quantum algorithms that are faster than classical algorithms for the same problem~\cite{add1,add2,rn6,rn7}. The continuous interest in quantum walk (QW) can be attributed to its broad applications to many distinct fields, such as polymer physics, solid state physics, biological physics, and quantum computation~\cite{rn6,rn7,rn1,rn2,rn3,rn4,rn5}. In the literature~\cite{rn1,rn2,rn3}, there are two types of quantum walks: continuous-time and discrete-time quantum walks. The main difference of the two types of quantum walks is that discrete-time quantum walks (DTQWs) require an extra coin Hilbert space in which the coin operator acts, while continuous-time quantum walks (CTQWs) do not need this extra Hilbert space. Aside from this, these two QWs are similar to their classical counterparts. Discrete-time quantum walks evolve by the application of a unitary evolution operator at discrete time intervals, and continuous-time quantum walks evolve under a (usually time-independent) Hamiltonian in Schr\"{o}dinger picture. Due to the different dimensional Hilbert space, CTQWs cannot be regarded as the limit of DTQWs as the time step goes to zero and there is no simple relation connecting the two QW models~\cite{rn8,childs,rn9}. However, in Ref.~\cite{childs}, the author proposes a precise correspondence between CTQWs and DTQWs on arbitrary graphs, showing that CTQWs can be obtained as an appropriate limit of DTQWs. The correspondence also leads to a new technique for simulating Hamiltonian dynamics, giving efficient simulations even in cases where the Hamiltonian is not sparse~\cite{childs}.

In this paper, we focus on the DTQWs. There are numerous studies on DTQWs on the line or cycle. However, all the studies for 1D DTQWs employ a moving shift operator $\hat{S}^m$, {\it i.e.}, $\hat{S}^m|j,L\rangle=|j-1,L\rangle$, $\hat{S}^m|j,R\rangle=|j+1,R\rangle$, the moving shift operator acting on a state only moves the position of the particle and does not change the direction of the coin. In the meanwhile, DTQWs using the swapping shift operator, which changes both the position and direction of the coin's quantum state, {\it i.e.}, $\hat{S}^s|j,L\rangle=|j-1,R\rangle$, $\hat{S}^s|j,R\rangle=|j+1,L\rangle$, have not received much attention in the literature. In Ref.~\cite{cycles}, the authors obtained the limiting probability
distributions of DTQWs using a Hadamard coin and moving shift operator on the cycle. There is no exact solutions for the limiting probability distribution of DTQWs on cycles using a general coin or swapping shift operator. In this paper, we will study DTQWs on cycles using a general coin and swapping shift operator,
and obtain exact solutions for the limiting probability distributions for the first time. In addition, based on the exact solutions, we analyze the symmetry behavior of the probability distributions. The symmetry analysis may be important for the controlling of QWs in experimental implementation. Before our findings, a well known feature is that the unbiased initial coin state $(|L\rangle\pm i |R\rangle)/\sqrt{2}$ leads to a symmetric probability distribution for the 1D quantum walks. This universal symmetry does not dependent on the coin parameters and holds for a wide range of quantum walks. The essential nature of such symmetry can be revealed by combining probabilities from two mirror image orthogonal components of the amplitudes. However, in addition to this universal symmetry, there are other initial coin states could result in a symmetric quantum walk. Here, we will determine a universal condition under which a symmetric quantum walk appears. Our results suggest that, in addition to the unbiased initial coin state $(|L\rangle\pm i |R\rangle)/\sqrt{2}$, other initial coin states can also realize a symmetric quantum walk.

\section{The Model}
In this section, we will define the model of discrete-time quantum walks on the cycles, and determine the eigenvalues and eigenstates of the evolution operator.
\subsection{Discrete-time quantum walks on the cycles (DTQWs): Initial state, Coins and Shift operator}
To address the problem, let's consider a 1D DTQW on the cycles. For a one-dimensional cycle composed of $N$ nodes, which are labeled as $\{x:x=1,2,...,N\}$, each node $1\leqslant x \leqslant N$ is connected its two nearest neighbors. The cycle is the simplest one-dimensional graph with periodic boundary condition. The Hilbert space ${\cal H}$ for DTQWs on cycles has $2N$ base vectors, which are denoted as $|1,L\rangle, |2,L\rangle, \cdots, |N,L\rangle, |1,R\rangle, |2,R\rangle, \cdots, |N,R\rangle$. Suppose the particle was initially ($t = 0$) localized at node $x_0$ and the initial coin states distributed in the coin subspace and superposed state $|C_0\rangle= p_0|L\rangle+e^{i\phi}q_0|R\rangle$ ($0<p_0,q_0<1$, $p_0^2+q_0^2=1, \phi\in [-\pi, \pi]$), {\it i.e.}, the initial state in the whole Hilbert space is,
\begin{equation}\label{eq01}
|\psi(0)\rangle= |x_0\rangle \otimes |C_0\rangle = |x_0\rangle  \otimes (p_0|L\rangle+e^{i\phi}q_0|R\rangle ) , 0<p_0,q_0<1, p_0^2+q_0^2=1, \phi\in [-\pi, \pi]
\end{equation}
where $\phi$ is the relative phase between the $|L\rangle$ and $|R\rangle$ amplitude.

For the coin operator, without loss of generality, we use the simple coin operator with one free parameter,
\begin{equation}\label{eq02}
\hat{C}=\begin{pmatrix}
       a &\ \ \ b \\
       b &\ \  -a
      \end{pmatrix}, \ 0<a,b<1, \ a^2+b^2=1.
\end{equation}

The initial state and coin operator in Eqs.~(\ref{eq01}) and (\ref{eq02}) are widely used in the theoretical models and experimental implementations of quantum walks~\cite{rn1,rn2,rn3}. For the controlling of evolution, we use the swapping shift operator $\hat{S}$, which changes both the position
and direction of the particle's quantum state, {\it i.e.},
\begin{equation}\label{eq03}
\begin{aligned}
\hat{S}|x,L\rangle=& |x-1,R\rangle    \\
\hat{S}|x,R\rangle=&|x+1,L\rangle
\end{aligned}
\end{equation}

The evolution of QW is governed by the evolution operator $\hat{U}=\hat{S}(\hat{C}\otimes \hat{I}_p) $ ($\hat{I}_p$ is the identity operator). The quantum state after $t$ steps is given by,
\begin{equation}\label{eq1}
|\psi(t)\rangle= \hat{U}^t |\psi(0)\rangle
\end{equation}
The probability of finding the particle at node $x$ after $t$ steps is,
\begin{equation}\label{eq2}
\begin{aligned}
P(x,t)&=|\langle x,L|\psi(t)\rangle|^2+|\langle x,R|\psi(t)\rangle|^2 \\
&= |\langle x,L|\hat{U}^t|\psi(0)\rangle|^2+ |\langle x,R|\hat{U}^t|\psi(0)\rangle|^2.
\end{aligned}
\end{equation}
Suppose the eigenvalue equation of $\hat{U}$ is $\hat{U}|\Psi_{j,J}\rangle=u_{j,J}|\Psi_{j,J}\rangle$ ($j\in[1,N], J\in \{+,-\}$), where $u_{j,J}$ and $|\Psi_{j,J}\rangle$ are the eigenvalues and orthonormalized eigenstates of the evolution operator $\hat{U}$. In the eigenstate space, the evolution operator is diagonalized as $\hat{U}^t= \sum_{j,J}u_{j,J}^t|\Psi_{j,J}\rangle\langle\Psi_{j,J}|$. Thus Eq.~(\ref{eq2}) can be written as,
\begin{small}
\begin{equation}\label{eq3}
\begin{aligned}
P(x,t)&=|\sum_{j,J} u_{j,J}^t \langle x,L|\Psi_{j,J}\rangle \langle \Psi_{j,J}|\psi(0)\rangle|^2   \\
&+ |\sum_{j,J} u_{j,J}^t \langle x,R|\Psi_{j,J}\rangle \langle \Psi_{j,J}|\psi(0)\rangle|^2 \\
&=\sum_{j,J}\sum_{j',J'}u_{j,J}^tu_{j',J'}^{*t}  \langle \Psi_{j,J}|\psi(0)\rangle \langle\psi(0)|\Psi_{j',J'}\rangle  \\
&(\langle x,L|\Psi_{j,J}\rangle \langle \Psi_{j',J'}|x,L\rangle + \langle x,R|\Psi_{j,J}\rangle \langle \Psi_{j',J'}|x,R\rangle)
\end{aligned}
\end{equation}
\end{small}
Noting that $\hat{U}$ is a unitary operator, {\it i.e.}, $\hat{U}\hat{U}^{\dag}=\hat{I}$, which leads to $|u|=1$ and $\lim_{T\rightarrow\infty}\frac{1}{T}\sum_{t=0}^T (u_{j,J}u_{j',J'}^*)^t =\delta(u_{j,J}-u_{j',J'})$,
the long time averages of $P(x,t)$ can be written as,
\begin{small}
\begin{equation}\label{eq4}
\begin{aligned}
\pi(x)&=\lim_{T\rightarrow\infty}\frac{1}{T}\sum_{t=0}^TP(x,t) \\
&=\sum_{j,J}\sum_{j',J'}\delta(u_{j,J}-u_{j',J'})  \langle \Psi_{j,J}|\psi(0)\rangle \langle\psi(0)|\Psi_{j',J'}\rangle \\
&\ \ ( \langle x,L|\Psi_{j,J}\rangle  \langle\Psi_{j',J'}|x,L\rangle+ \langle x,R|\Psi_{j,J}\rangle  \langle\Psi_{j',J'}|x,R\rangle )
\end{aligned}
\end{equation}
\end{small}
where $\delta(u_{j,J}-u_{j',J'})$ takes value 1 if $u_{j,J}=u_{j',J'}$ and equals to 0 otherwise. In the above equation, we can see that the limit distribution $\pi(x)$ depends on the eigenvalues and eigenstates of the evolution operator $\hat{U}$. The limiting probability $\pi(x)$ in Eq.~(\ref{eq4}) is also called stationary probability, which reflecting the equilibrium of the system evolution. In order to calculate the analytical expressions for $P(x,t)$ and $\pi(x)$, all the eigenvalues $u_{j,J}$ and eigenstates $|\Psi_{j,J}\rangle$ of the evolution operator $\hat{U}$ are required.

\subsection{Eigenvalues and eigenstates of the evolution operator $\hat{U}$}
In the Appendix, we use the technique of Chebyshev polynomials to calculate the $2N$ eigenvalues of evolution operator $\hat{U}$ as follows (See Eq.~(\ref{a15}) in the Appendix),
\begin{equation}\label{eq5}
\begin{aligned}
u_{j,\pm}&=b\cos\theta_j \pm i\sqrt{1-b^2\cos^2\theta_j}, \theta_j=\frac{2j\pi}{N}, j\in [1,N].
\end{aligned}
\end{equation}
We also determine the $2N$ orthonormalized and normalized eigenstates of $\hat{U}$, which can be expanded in the Bloch states as (See Eqs.~(\ref{a17})-(\ref{a19}) in the Appendix),
\begin{equation}\label{eq6}  \small
\begin{aligned}
|\Psi&_{j,\pm}\rangle=\sum_{k=1}^N  \frac{ae^{ik\theta_j}}{\sqrt{N[a^2+(\sqrt{1-b^2\cos^2\theta_j} \pm b\sin\theta_j)^2}]}  \\
& \Big [|k,L\rangle \mp \frac{i}{a} e^{i\theta_j}\big(\sqrt{1-b^2\cos^2\theta_j} \pm b\sin\theta_j \big) |k,R\rangle \Big]
\end{aligned}
\end{equation}
For the convenient of calculation, we write Eq.~(\ref{eq6}) into the following simple form,
\begin{equation}\label{eq60}
\begin{aligned}
|\Psi_{j,\pm}\rangle = & \sum_{k=1}^N Z_{j,\pm}e^{ik\theta_j} \Big (|k,L\rangle + M_{j,\pm}e^{i\theta_j\mp \frac{i\pi}{2}} |k,R\rangle \Big), \\
\text{where}\ Z_{j,\pm}=& \frac{a}{\sqrt{N[a^2+(\sqrt{1-b^2\cos^2\theta_j} \pm b\sin\theta_j)^2}]} \\
\text{and}\   M_{j,\pm}=& \frac{1}{a} \big(\sqrt{1-b^2\cos^2\theta_j} \pm b\sin\theta_j \big).
\end{aligned}
\end{equation}
After some algebra calculation, we find several useful identities between $Z_{j,\pm}$ and $M_{j,\pm}$,
\begin{eqnarray}
Z_{j,\pm}^2=\frac{\sqrt{1-b^2\cos^2\theta_j} \mp b\sin\theta_j}{2N\sqrt{1-b^2\cos^2\theta_j}},  \label{eq61} \\
Z_{j,\pm}^2M_{j,\pm}^2=\frac{\sqrt{1-b^2\cos^2\theta_j} \pm b\sin\theta_j}{2N\sqrt{1-b^2\cos^2\theta_j}}   , \label{eq62}\\
Z_{j,+}Z_{j,-}=Z_{j,\pm}^2M_{j,\pm} = \frac{a}{2N\sqrt{1-b^2\cos^2\theta_j}}    , \label{eq63} \\
Z_{j,\pm}^2(1+M_{j,\pm}^2)=\frac{1}{N}, \label{eq64}   \\
M_{j,+}M_{j,-}=1, \label{eq65} \\
Z_{j,+}M_{j,+}=Z_{j,-};\ \  Z_{j,-}M_{j,-}=Z_{j,+} \label{eq66}
\end{eqnarray}

\section{Results}
In this section, we will use the eigenstates and identities in Eqs.~(\ref{eq60})-(\ref{eq66}) to analyze the limiting probability distributions.
\subsection{Limiting probability distributions}
According to Eq.~(\ref{eq4}), the long-time averaged distribution depends on the eigenstates $|\Psi_{j,\pm}\rangle$ and degeneracy of eigenvalues.
Eq.~(\ref{eq5}) suggests that most of the eigenvalues are double-fold degenerate. Concretely, if the cycle size N is an even number, there are two nondegenerate eigenvalues ($j=N/2, N$), the other eigenvalues have degeneracy 2 ($j'=N-j$, $u_{j,\pm}=u_{N-j,\pm}$). If the cycle size N is an odd number, there is one nondegenerate eigenvalue ($j=N$), and the other eigenvalues have degeneracy 2 ($j'=N-j$, $u_{j,\pm}=u_{N-j,\pm}$). The limiting probability distribution in Eq.~(\ref{eq4}) can be divided into two parts: contribution from nondegenerate eigenvalues ($j=j', \ J=J'$) and contribution from degenerate eigenvalues ($j'=N-j,\ J=J'$).
In Eq.~(\ref{eq4}), the summation over $J$ involves the contributions from eigenstates with different $J=\{+,-\}$ sign. For the sake of simplicity, the summation over $J$ is always indicated/included by the subscript $\pm$ and the notation of $\sum_J$ is omitted in the following. Consequently, according to Eq.~(\ref{eq4}), the contributions from nondegenerate eigenvalues ($j=j', \ J=J'$) can be easily written as,
\begin{equation}\label{eq7} \small
S_1= \sum_{j,\pm}   |\langle \psi(0) |\Psi_{j,\pm}\rangle|^2  \big ( |\langle x,L|\Psi_{j,\pm}\rangle|^2+|\langle x,R|\Psi_{j,\pm}\rangle|^2 \big)
\end{equation}
Likewise, the contributions from degenerate eigenvalues ($j'=N-j$) can be recasted as,
\begin{equation}\label{eq8} \small
\begin{aligned}
&S_2= \sum_{j=1,\pm}^{N-1} \langle \Psi_{j,\pm}|\psi(0)\rangle \langle\psi(0)|\Psi_{N-j,\pm}\rangle \\
& \big(\langle x,L|\Psi_{j,\pm}\rangle \langle \Psi_{N-j,\pm}|x,L\rangle + \langle x,R|\Psi_{j,\pm}\rangle \langle \Psi_{N-j,\pm}|x,R\rangle \big)
\end{aligned}
\end{equation}
If the cycle size N is an even number, the contributions of $j=\frac{N}{2}$ in Eqs.~(\ref{eq7}) and (\ref{eq8}) are the same. When adding $S_1$ and $S_2$, the contributions of $j=\frac{N}{2}$ are calculated twice. In this case, we need to deduct the contribution of $j=\frac{N}{2}$, which is given by,
\begin{equation}\label{eq9}  \small
S_3= \sum_{\pm} |\langle\psi(0)|\Psi_{\frac{N}{2},\pm}\rangle|^2
 \Big( |\langle x,L|\Psi_{\frac{N}{2},\pm}\rangle|^2+  |\langle x,R|\Psi_{\frac{N}{2},\pm}\rangle|^2 \Big)
\end{equation}
Thus the limiting probability in Eq.~(\ref{eq4}) can be written as,
\begin{equation}\label{eq10}
\pi(x)= S_1 + S_2 -\delta_{mod(N,2),0} S_3,
\end{equation}
where $\delta_{mod(N,2),0}$ equals to 1 for even-numbered N and 0 otherwise.

Now we use the initial state in Eq.~(\ref{eq01}) and eigenstates in Eq.~(\ref{eq60}) to calculate $S_1$, $S_2$ and $S_3$. Substituting $|\psi(0)\rangle$ and $|\Psi_{j,\pm}\rangle$ in Eq.~(\ref{eq60}) into the Eq.~(\ref{eq7}), we find that $|\langle x,L|\Psi_{j,\pm}\rangle|^2+|\langle x,R|\Psi_{j,\pm}\rangle|^2= Z_{j,\pm}^2(1+M_{j,\pm}^2)=\frac{1}{N}$ (Identity (\ref{eq64}) has been applied). $|\langle \psi(0) |\Psi_{j,\pm}\rangle|^2$ is simplified as,
\begin{equation}\label{eq11}
\begin{aligned}
|\langle \psi(0) |\Psi_{j,\pm}\rangle|^2 =& |p_0Z_{j,\pm} + q_0M_{j,\pm}Z_{j,\pm}e^{i\theta_j-i\phi \mp\frac{i\pi}{2}}|^2  \\
=& p_0^2Z_{j,\pm}^2 + q_0^2Z_{j,\pm}^2M_{j,\pm}^2  +  2p_0q_0 Z_{j,\pm}^2M_{j,\pm}\cos(\theta_j-\phi \mp\frac{\pi}{2}) \\
=&p_0^2\cdot\frac{\sqrt{1-b^2\cos^2\theta_j} \mp b\sin\theta_j}{2N\sqrt{1-b^2\cos^2\theta_j}}+ q_0^2\cdot \frac{\sqrt{1-b^2\cos^2\theta_j} \pm b\sin\theta_j}{2N\sqrt{1-b^2\cos^2\theta_j}}  \\
& \pm 2p_0q_0 Z_{j,\pm}^2M_{j,\pm}\sin(\theta_j-\phi) \\
=&\frac{1}{2N}\pm \frac{b(q_0^2-p_0^2)\sin\theta_j}{2N\sqrt{1-b^2\cos^2\theta_j}} \pm \frac{ap_0q_0\sin(\theta_j-\phi)}{N\sqrt{1-b^2\cos^2\theta_j}}
\end{aligned}
\end{equation}
where identities (\ref{eq61}), (\ref{eq62}) and (\ref{eq63}) have been used in the above calculation.
Since the second and third terms in the above equation are odd function of $\theta_j$, the summation over $j$ for the last two terms equals to zero respectively. Only the first term in Eq.~(\ref{eq11}) gives essential contribution to Eq.~(\ref{eq7}), which leads to $S_1=1/N$. Setting $j=N/2$, $\theta_j=\pi$, the second and third terms in Eq.~(\ref{eq11}) equal to zero, which leads to $S_3=1/N^2$ in Eq.~(\ref{eq9}).

Next we calculate $S_2$ in Eq.~(\ref{eq8}). The two product terms in Eq.~(\ref{eq8}) are related to the final position $x$ and starting position $x_0$.
Noting that  $\langle x,L|\Psi_{j,\pm}\rangle= Z_{j,\pm}e^{ix\theta_j}$, $\langle x,R|\Psi_{j,\pm}\rangle= Z_{j,\pm}M_{j,\pm}e^{ix\theta_j}e^{i\theta_j\mp\frac{i\pi}{2}}$, $\langle \Psi_{N-j,\pm}|x,L\rangle = Z_{j,\mp}e^{ix\theta_j}$ and $\langle \Psi_{N-j,\pm}|x,R\rangle= Z_{j,\mp}M_{j,\mp}e^{ix\theta_j}e^{i\theta_j\pm\frac{i\pi}{2}}$, we obtain
\begin{equation}\label{eq12}
\begin{aligned}
&\langle x,L|\Psi_{j,\pm}\rangle \langle \Psi_{N-j,\pm}|x,L\rangle + \langle x,R|\Psi_{j,\pm}\rangle \langle \Psi_{N-j,\pm}|x,R\rangle  \\
&=Z_{j,\pm}Z_{j,\mp}e^{2ix\theta_j} + Z_{j,\pm}Z_{j,\mp}M_{j,\pm}M_{j,\mp}e^{2ix\theta_j+2i\theta_j} \\
&=Z_{j,\pm}Z_{j,\mp}e^{2ix\theta_j} (1+ M_{j,\pm}M_{j,\mp}e^{2i\theta_j} )     \\
&=\frac{ae^{2ix\theta_j}(1+e^{2i\theta_j})}{2N\sqrt{1-b^2\cos^2\theta_j}}.  \ \ \  \text{Identities (\ref{eq63}) and (\ref{eq65})}  \  \text{are used}.
\end{aligned}
\end{equation}
Substituting Eq.~(\ref{eq01}) into Eq.~(\ref{eq8}), we arrive at $\langle \Psi_{j,\pm}|\psi(0)\rangle=e^{-ix_0\theta_j}(p_0Z_{j,\pm} + q_0Z_{j,\pm}M_{j,\pm}e^{-i\theta_j+i\phi\pm\frac{i\pi}{2}})$ and $\langle\psi(0)|\Psi_{N-j,\pm}\rangle= e^{-ix_0\theta_j}(p_0Z_{j,\mp} + q_0Z_{j,\mp}M_{j,\mp}e^{-i\theta_j-i\phi\mp\frac{i\pi}{2}})$. The term $\langle \Psi_{j,\pm}|\psi(0)\rangle \langle\psi(0)|\Psi_{N-j,\pm}\rangle$ becomes
\begin{equation}\label{eq13}
\begin{aligned}
&\langle \Psi_{j,\pm}|\psi(0)\rangle \langle\psi(0)|\Psi_{N-j,\pm}\rangle =e^{-2ix_0\theta_j}(p_0Z_{j,\pm} + q_0Z_{j,\pm}M_{j,\pm}e^{-i\theta_j+i\phi\pm\frac{i\pi}{2}})   \\
&\times (p_0Z_{j,\mp} + q_0Z_{j,\mp}M_{j,\mp}e^{-i\theta_j-i\phi\mp\frac{i\pi}{2}}) \\
&= e^{-2ix_0\theta_j}\Big( p_0^2Z_{j,\pm}Z_{j,\mp} + q_0^2 Z_{j,\pm}Z_{j,\mp}M_{j,\pm}M_{j,\mp}e^{-2i\theta_j} \\
& + p_0q_0Z_{j,\pm}Z_{j,\mp}M_{j,\pm}e^{-i\theta_j+i\phi\pm\frac{i\pi}{2}} + p_0q_0Z_{j,\pm}Z_{j,\mp}M_{j,\mp}e^{-i\theta_j-i\phi\mp\frac{i\pi}{2}} \Big) \\
&=e^{-2ix_0\theta_j}Z_{j,\pm}Z_{j,\mp}\Big[ p_0^2 + q_0^2 M_{j,\pm}M_{j,\mp}e^{-2i\theta_j} + p_0q_0e^{-i\theta_j}(M_{j,\pm}e^{i\phi\pm\frac{i\pi}{2}} + M_{j,\mp}e^{-i\phi\mp\frac{i\pi}{2}}) \Big] \\
&=\frac{ae^{-2ix_0\theta_j}}{2N\sqrt{1-b^2\cos^2\theta_j}} \cdot \Big[ p_0^2+q_0^2e^{-2i\theta_j} + p_0q_0e^{-i\theta_j}\big(  \frac{\sqrt{1-b^2\cos^2\theta_j} \pm b\sin\theta_j}{a}e^{i\phi\pm\frac{i\pi}{2}} \\
& + \frac{\sqrt{1-b^2\cos^2\theta_j} \mp b\sin\theta_j}{a}e^{-i\phi\mp\frac{i\pi}{2}}        \big) \Big].\ \ \text{Identities (\ref{eq63}) and (\ref{eq65})}  \  \text{are used}.
\end{aligned}
\end{equation}
In Eq.~(\ref{eq13}), the first two terms ($p_0^2$ and $q_0^2e^{-2i\theta_j}$) in the bracket do not depend on the parity sign $\pm$, the summation over $\pm$ will be double. In contrast, the last two terms in the small bracket $()$ depend on the parity sign $\pm$, and the summation over $\pm$ leads to $\sum_{\pm}(M_{j,\pm}e^{i\phi\pm\frac{i\pi}{2}} + M_{j,\mp}e^{-i\phi\mp\frac{i\pi}{2}})=\frac{2ib\sin\theta_j}{a}(e^{i\phi}+e^{-i\phi})=\frac{2b\cos\phi}{a}(e^{i\theta_j}-e^{-i\theta_j})$. Finally, Eq.~(\ref{eq13}) is simplified as,
\begin{equation}\label{eq14}
\begin{aligned}
&\sum_{\pm}\langle \Psi_{j,\pm}|\psi(0)\rangle \langle\psi(0)|\Psi_{N-j,\pm}\rangle =\frac{ae^{-2ix_0\theta_j}}{N\sqrt{1-b^2\cos^2\theta_j}}
\cdot \Big[ (p_0^2+\frac{bp_0q_0\cos\phi}{a})   +  (q_0^2 - \frac{bp_0q_0\cos\phi}{a})e^{-2i\theta_j} \Big].
\end{aligned}
\end{equation}
Combining Eqs.~(\ref{eq12}) and (\ref{eq14}), we obtain a simple form for $S_2$
\begin{equation}\label{eq15}
\begin{aligned}
&S_2=\frac{a^2}{2N^2} \sum_{j=1}^{N-1}  \frac{e^{2i(x-x_0)\theta_j}}{1-b^2\cos^2\theta_j} \cdot
 \Big[ 1+ (p_0^2 + \frac{bp_0q_0\cos\phi}{a})e^{2i\theta_j}   +  (q_0^2 - \frac{bp_0q_0\cos\phi}{a})e^{-2i\theta_j} \Big],
\end{aligned}
\end{equation}
which is a function of the distance $d\equiv x-x_0$ between $x$ and $x_0$. Noting that $S_1=1/N$ and $S_3=1/N^2$, the limiting probability distribution $\pi(x)\equiv\pi(d)$ is closely related to $S_2$. Here we obtain exact solutions for the limiting probability distribution $\pi(x)$ in Eq.~(\ref{eq10}),
\begin{equation}\label{eq150}
\begin{aligned}
\pi(d)=&\pi(x-x_0)=   \frac{1}{N}-\delta_{mod(N,2),0}\frac{1}{N^2} + \\
&\frac{a^2}{2N^2} \sum_{j=1}^{N-1}  \frac{e^{2i(x-x_0)\theta_j}}{1-b^2\cos^2\theta_j} \cdot
 \Big[ 1+ (p_0^2 + \frac{bp_0q_0\cos\phi}{a})e^{2i\theta_j}   +  (q_0^2 - \frac{bp_0q_0\cos\phi}{a})e^{-2i\theta_j} \Big].
\end{aligned}
\end{equation}
which is crucial to analyze the symmetry of quantum walks. Here for the first time, we obtain the exact solutions for the limiting probability distribution for QWs using a general coin and swapping shift operator.

\subsection{Symmetry analysis and mixing time}
In the following, we use the exact solution of the limiting probability distribution $\pi(d)$ to determine a general condition under which the quantum walk is symmetric. The symmetry of the limiting probability distribution requires $\pi(d)=\pi(-d)$. The summation in Eq.~(\ref{eq15}) is a real values, the imaginary part vanished when summing over $j$. Thus $S_2$ can be rewritten as the summation of the real part of the terms,
\begin{equation} \label{eq151}
\begin{aligned}
S_2(d)=&\frac{a^2}{2N^2} \sum_{j=1}^{N-1}  \frac{1}{1-b^2\cos^2\theta_j} \cdot
 \Big[ \cos2d\theta_j + (p_0^2 + \frac{bp_0q_0\cos\phi}{a})\cos2(d+1)\theta_j  \\
  &+  (q_0^2 - \frac{bp_0q_0\cos\phi}{a})\cos2(d-1)\theta_j \Big],
\end{aligned}
\end{equation}
In the above Equation, change $d$ to $-d$ arriving at,
\begin{equation} \label{eq152}
\begin{aligned}
S_2(-d)=&\frac{a^2}{2N^2} \sum_{j=1}^{N-1}  \frac{1}{1-b^2\cos^2\theta_j} \cdot
 \Big[ \cos2d\theta_j + (p_0^2 + \frac{bp_0q_0\cos\phi}{a})\cos2(d-1)\theta_j  \\
  &+  (q_0^2 - \frac{bp_0q_0\cos\phi}{a})\cos2(d+1)\theta_j \Big],
\end{aligned}
\end{equation}

\begin{figure}
\scalebox{0.5}[0.5]{\includegraphics{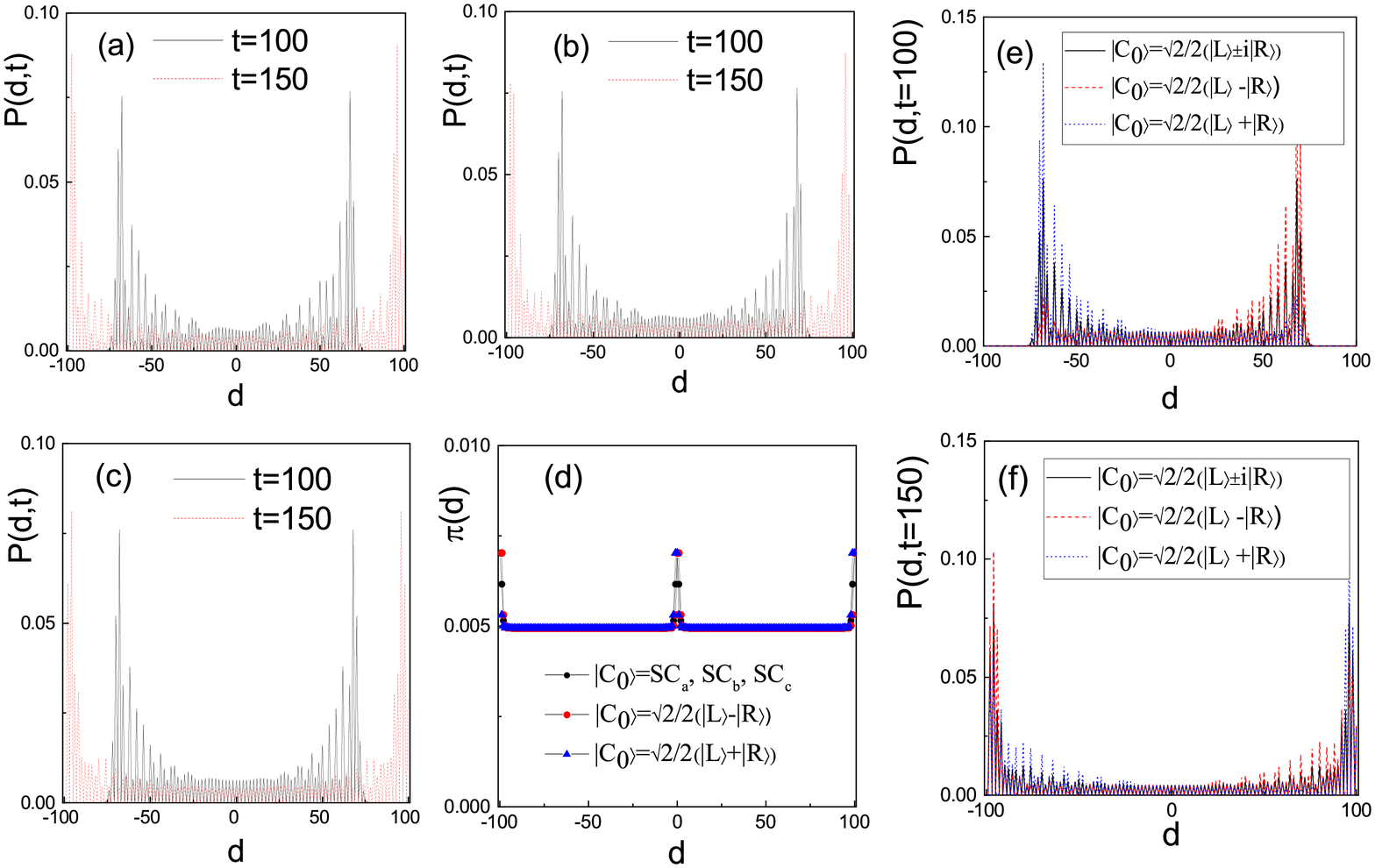}}
\caption{Probability distribution $P(d,t)$ for Hadamard quantum walks ($a=b=\sqrt{2}/2$) on cycles of size $N=200$ after $t=100$ (black curves) and $t=150$ (dashed red curves) steps for three different initial coin states: ($CS_a$) initial coin state $|C_0\rangle=\sqrt{\frac{2-\sqrt{2}}{4}}|L\rangle +\sqrt{\frac{2+\sqrt{2}}{4}}|R\rangle$, ($CS_b$) initial coin state $|C_0\rangle=\sqrt{\frac{5-\sqrt{5}}{10}}|L\rangle +\sqrt{\frac{5+\sqrt{5}}{10}}e^{\pm\frac{i\pi}{3}}|R\rangle$ and ($CS_c$) unbiased initial coin state  $|C_0\rangle=\frac{\sqrt{2}}{2}(|L\rangle \pm i|R\rangle) $. (d) is the limiting probability distributions $\pi(d)$ for the three different initial coin states ($CS_a$), ($CS_b$) and ($CS_c$), which are exactly the same(see black squares), as well as the biased initial coin states $|C_0\rangle=\frac{\sqrt{2}}{2}(|L\rangle -|R\rangle$ (see the red dots) and $|C_0\rangle=\frac{\sqrt{2}}{2}(|L\rangle +|R\rangle$ (see the blue triangles). (e) is the symmetric and asymmetric probability distributions $P(d,t)$ after $t=100$ steps for Hadamard quantum walks using the initial coin states $|C_0\rangle=\frac{\sqrt{2}}{2}(|L\rangle \pm i|R\rangle) $ (black solid curve, symmetric distribution), $|C_0\rangle=\frac{\sqrt{2}}{2}(|L\rangle-|R\rangle) $ (red dashed curve, asymmetric distribution) and $|C_0\rangle=\frac{\sqrt{2}}{2}(|L\rangle -|R\rangle) $ (blue dotted curve, asymmetric distribution). (f) is the similar probability distributions for $P(d,t)$ after $t=150$ steps. The initial coin states $CS_a$, $CS_b$ and $CS_c$ satisfied Eq.~(\ref{eq16}), thus the probability distributions shown in (a)-(c) are nearly the same. \label{fig1}}
\end{figure}

If $S_2(d)=S_2(-d)$, coefficients in the parentheses are equal, which lead to,
\begin{equation}\label{eq16}
2\frac{b}{a}p_0q_0\cos\phi =q_0^2-p_0^2.
\end{equation}
The above equation is the condition under which the quantum walk is symmetric, one of the main conclusions of this paper. It is evident that the unbiased initial coin state $p_0=q_0=\sqrt{2}/2,\ \phi=\pm\pi/2$ satisfy the above condition. In addition to this unbiased initial coin state, there are other solutions for Eq.~(\ref{eq16}). In the literature, it is well known that the unbiased initial coin state $(|L\rangle\pm i |R\rangle)/\sqrt{2}$ leads to a symmetric probability distribution for the 1D quantum walks. Here we show that, in addition to this universal symmetry, there are other initial coin states could result in a symmetric distribution. In order to compare the symmetric behavior, we choose two additional solutions for further study. For the Hadamard walk $a=b=\sqrt{2}/2$, we choose initial coin states: ($CS_a$) $p_0=\sqrt{\frac{2-\sqrt{2}}{4}}=\sin\frac{\pi}{8}, q_0=\sqrt{\frac{2+\sqrt{2}}{4}}=\cos\frac{\pi}{8}, \phi=0$ and ($CS_b$) $p_0=\sqrt{\frac{5-\sqrt{5}}{10}}, q_0=\sqrt{\frac{5+\sqrt{5}}{10}}, \phi=\pm\frac{\pi}{3}$, as well as the unbiased initial coin state ($CS_c$) $p_0=q_0=\sqrt{2}/2,\ \phi=\pm\pi/2$ for comparison.  Fig.~\ref{fig1} (d) shows the limiting probability distributions of the Hadamard walks with the three different initial coin states, which are exactly the same and satisfy $\pi(d)=\pi(-d)$ (See the black squares in Fig.~\ref{fig1} (d)).

It is worth mentioning that if the initial state and coin parameters satisfy Eq.~(\ref{eq16}), the limiting probability distribution is symmetric. Now a natural question is that whether the evolution probability $P(d=x-x_0,t)$ is also symmetric during all the time. To address this question, we show the evolution probability $P(d,t)$ for Hadamard quantum walks with the three different initial states in Fig.~\ref{fig1}(a)-(c),(e)-(f). We can see that the evolution probability distributions are nearly the same, and the three different initial states give almost identical probability distributions. The symmetry for the unbiased initial coin state exactly satisfy $P(d,t)=P(-d,t)$ while the other initial states do not have such strict symmetry. This feature is consistent with the results in Ref.~\cite{rn11} where the initial state $|C_0\rangle=\sin\frac{\pi}{8}|L\rangle + \cos\frac{\pi}{8}|R\rangle$ nearly leads to a symmetric probability distribution. It is obvious that there are two distinct ways of arriving at a symmetric quantum walk, one is obtained by combining probabilities from two mirror image orthogonal components ($p_0=q_0=\sqrt{2}/2, \phi=\pm\pi/2$), the others are obtained by interference.
\begin{figure}
\scalebox{0.25}[0.25]{\includegraphics{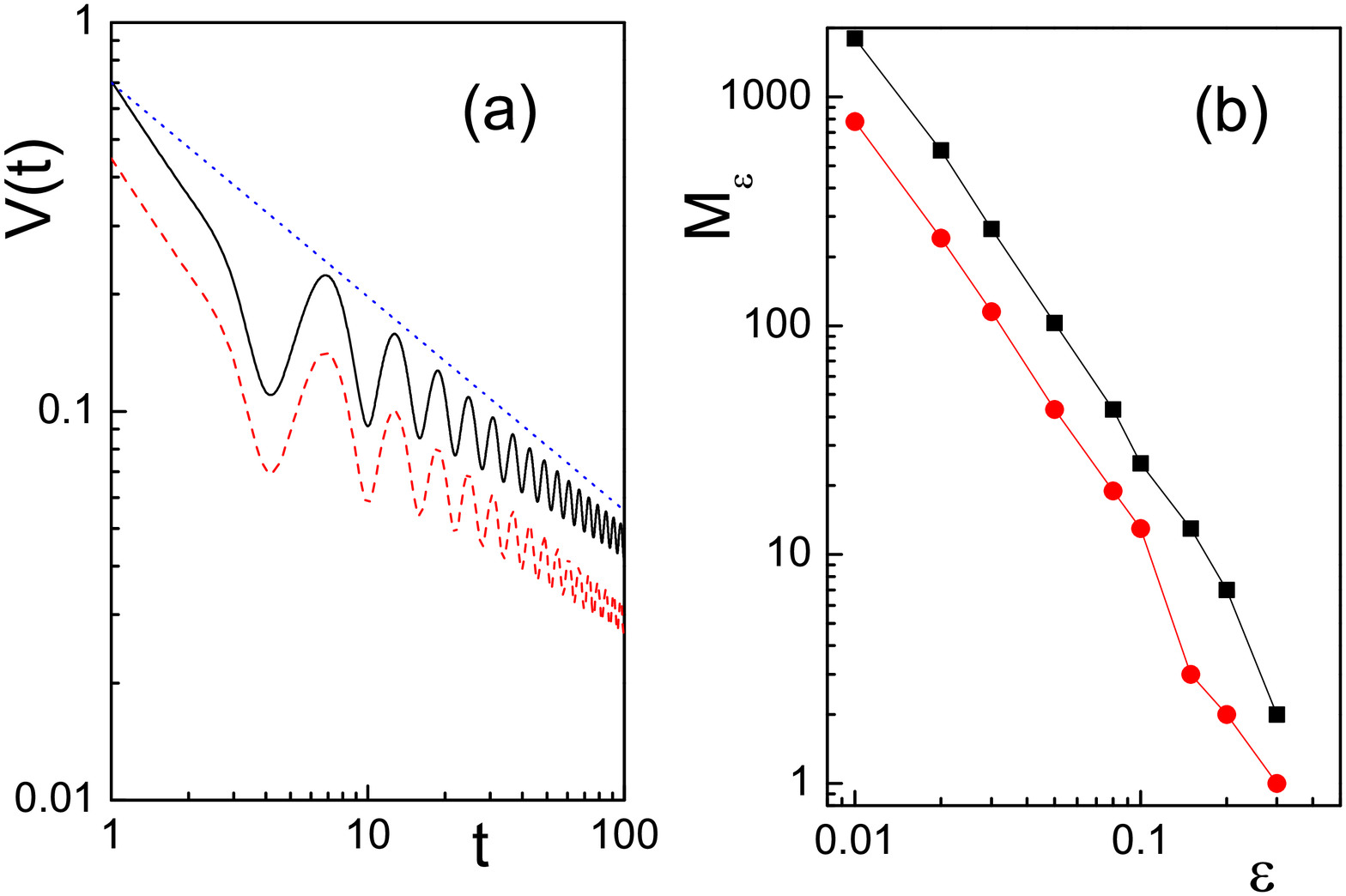}} \caption{
(Color online) (a) Dependence of the symmetry variation $V(t)$ on the evolution time $t$ for Hadamard walks with initial coin states ($CS_a$) $|C_0\rangle=\sqrt{\frac{2-\sqrt{2}}{4}}|L\rangle +\sqrt{\frac{2+\sqrt{2}}{4}}|R\rangle$ (black solid curve) and ($CS_b$) $|C_0\rangle=\sqrt{\frac{5-\sqrt{5}}{10}}|L\rangle +\sqrt{\frac{5+\sqrt{5}}{10}}e^{\pm\frac{i\pi}{3}}|R\rangle$ (red dashed curve).
The blue dotted line indicates the power law decay $t^{-0.5}$. (b) Dependence of the mixing time $M_\epsilon$ on the threshold value $\epsilon$ for Hadamard walks with the initial coin states ($CS_a$, black squares) and ($CS_b$, red dots). \label{fig2}}
\end{figure}
Except for the unbiased initial coin state $p_0=q_0=\sqrt{2}/2,\ \phi=\pm\pi/2$, the other initial coin states satisfying Eq.~(\ref{eq16}) only gives a proximate symmetric distribution of $P(d,t)$ while the symmetry of $\pi(d)$ is perfect. In order to make a quantitative analysis of the symmetry of the evolution probability, we use the variation $V(t)=\sum_d|P(d,t)-P(-d,t)|^2$ to measure the symmetry of the distribution. The smaller the $V(t)$ value, the more symmetric the quantum walk is. Fig.~\ref{fig2}(a) shows the time dependence of $V(t)$ for Hadamard walks with the initial coin states ($CS_a$) and ($CS_b$). As we can see, the symmetry variation $V(t)$ oscillate frequently and decays as a power law of $t^{-0.5}$. This suggests that the evolution probability of quantum walk converges to a symmetry distribution rapidly as the evolution time increased.

To quantify how fast the evolution probability converges to the symmetric distribution, we define a mixing time for the symmetry variation $M_{\epsilon}=min\{\tau \ |\ \forall \  t>\tau,\ V(t)< \epsilon\}$. Fig.~\ref{fig2}(b) shows the dependence of the mixing time $M_\epsilon$ on the threshold value $\epsilon$ for Hadamard walks with the initial coin states ($CS_a$) and ($CS_b$). As we can see, the mixing time also shows a power-law decay of $\epsilon$. This result is similar to mixing time behavior for quantum walks on Hypercube~\cite{rn12,rn13}. We also try to compare the symmetry variation $V(t)$ and mixing time $M_{\epsilon}$ for the other initial coin parameters and initial states. We find that quantum walk with the unbiased initial states has the smallest symmetry variation $V(t)$ and mixing time $M_{\epsilon}$. The Hadamard quantum walks have a smaller $V(t)$ than the other walks using biased coins ($a,b\neq\sqrt{2}/2$). This may suggest that quantum walks with unbiased coin parameters and initial states mix to the symmetric distribution fast. We hope this conclusion can be used in constructing efficient quantum algorithms.

\section{Conclusions}
In summary, we obtain exact analytical solutions of the long-time averaged probabilities for the 1D quantum walks for the first time. According to the analytical solutions, we determine a general condition under which the quantum walks are symmetric.  We show that, in addition to the symmetric initial coin state $(|L\rangle \pm i |R\rangle)/\sqrt{2}$ could lead to a symmetric probability distribution, choosing other appropriate initial state parameters ($p_0$, $q_0$, $\phi$) could also achieve a symmetric quantum walk. We define a symmetry variation $V(t)$ to quantify the symmetry and find that the evolution probability distribution converges to the symmetric distribution quickly. We hope such symmetric condition for quantum walks could provide useful insights in construction of efficient quantum algorithms.

{\it Acknowledgments:} This work is supported by the National Natural Science Foundation of China under project 11205110, Innovation and entrepreneurship training program for College Students under project 2015xj070. Yusuke Ide is supported by Yokohama Academic Foundation.

\appendix
\section{The matrix form of the evolution operator $\hat{U}$}
The swapping shift operator $\hat{S}$ swaps the particle's state, moving the particle to the neighboring position and changing the direction.
The swapping shift operator $\hat{S}$ acting on an arbitrary state$|i,J\rangle$ is summarized as,
\begin{equation}\label{a1}
\hat{S}|i,J\rangle=
\begin{cases}
|i+1,L\rangle,  & \text{if } J=R \\
|i-1,R\rangle,  & \text{if } J=L
\end{cases}
\end{equation}
The periodic boundary condition of the cycle requires $\hat{S}|1,L\rangle=|N,R\rangle$, $\hat{S}|N,R\rangle=|1,L\rangle$. The elements of the swapping shift operator $\hat{S}$ in the Hilbert space is,
\begin{equation}\label{a2}
\begin{aligned}
\langle i,J|\hat{S}|i',J'\rangle=
\begin{cases}
\delta_{i,i'+1},  & \text{if } J=L, J'=R \\
\delta_{i+1,i'},  & \text{if } J=R, J'=L \\
0, & \text{Otherwise.}
\end{cases}
\end{aligned}
\end{equation}

Noting that the coin operator $\hat{C}=\begin{pmatrix}
       a &\ \ \ b \\
       b &\ \  -a
      \end{pmatrix}$ and the relationship $\hat{U}=\hat{S}(\hat{I}_p\otimes \hat{C})$, we get the matrix form for the evolution operator $\hat{U}$

\begin{footnotesize}
\begin{equation}\label{a3}
\hat{U}=
\kbordermatrix{\mbox{}&1&\cdots&{\scriptscriptstyle N-1}&{\scriptscriptstyle N}&\vrule &1&\cdots&{\scriptscriptstyle N-1}&{\scriptscriptstyle N}\\
\cline{2-10}
1                        &0         &       &0          &b      &\vrule &0      &          &0       &-a    \\
2                        &b         &0      &           &       &\vrule &-a     &0         &        &         \\
\vdots                   &          &\ddots &0          &       &\vrule &       &\ddots    &0       &      \\
{\scriptscriptstyle N}   &          &       &b          &0      &\vrule &       &          &-a       &0     \\
\hline
1                        &0         &a      &           &       &\vrule &0      &b         &       &       \\
\vdots                   &          &0      &\ddots     &       &\vrule &       &0         &\ddots &     \\
{\scriptscriptstyle N-1} &          &       &0          &a      &\vrule &       &          &0      &b      \\
{\scriptscriptstyle N}   &a         &       &           &0      &\vrule &b       &          &       &0
}
\end{equation}
\end{footnotesize}
\section{Eigenvalues and eigenstates of the evolution operator $\hat{U}$}
In this section, we determine the eigenvalues and eigenstates for the evolution operator $\hat{U}$ obtained in Eq.~(\ref{a3}).
\subsection{Eigenvalues}
To obtain the eigenvalues of $\hat{U}$, we start our analysis on the eigenequation of the evolution operator $\hat{U}$ (see Eq.~(\ref{a3})). Suppose the eigenequation of $\hat{U}$ is $\hat{U}|\Psi\rangle=u|\Psi\rangle$ ($\hat{U}|\Psi_{j,J}\rangle=u_{j,J}|\Psi_{j,J}\rangle$), the eigenstates $|\Psi\rangle$ can be expanded as
\begin{equation}\small    \label{a4}
|\Psi\rangle=\sum_{j,J}\alpha_{j,J}|j,J\rangle=\sum_{j}\alpha_{j,L}|j,L\rangle+\sum_{j}\alpha_{j,R} |j,R\rangle
\end{equation}

Noting that the matrix form of $\hat{U}$ in Eq.~(\ref{a3}), the eigen equation $\hat{U}|\Psi\rangle=u|\Psi\rangle$ can be decomposed into the following $2N$ linear equations,
\begin{align}
b\alpha_{N,L}-a\alpha_{N,R} &=u\alpha_{1,L},    \label{a5} \\
b\alpha_{j,L}-a\alpha_{j,R} &=u\alpha_{j+1,L},  \ \ \ 1\leqslant j\leqslant N-1 \label{a6} \\
a\alpha_{j,L}+b\alpha_{j,R} &=u\alpha_{j-1,R},  \ \ \ 2\leqslant j\leqslant N \label{a7} \\
a\alpha_{1,L}+b\alpha_{1,R} &=u\alpha_{N,R}, \label{a8}
\end{align}
Utilizing Eq.~(\ref{a6}) to eliminate $\alpha_{j,R}$ and $\alpha_{j-1,R}$, Eq.~(\ref{a7}) becomes $\alpha_{j+1,L}+\alpha_{j-1,L}=\frac{u^2+1}{bu}\alpha_{j,L}$. This is similar to the recursive relation of the Chebyshev
polynomials (see Appendix A in Ref.~\cite{rn14}). Noting the recursive relations and the mapping relationship $\frac{u^2+1}{bu}\equiv 2x$ in the definition
of the Chebyshev polynomials of the second kind, the variables $\alpha_{j,L}$ ($j\in [1,N]$) can be expressed as a function of $\alpha_{2,L}$ and $\alpha_{1,L}$,
\begin{equation}\label{a9}
\alpha_{j,L}=U_{j-2}(x)\alpha_{2,L}-U_{j-3}(x)\alpha_{1,L}, \ \ j\in [1,N]
\end{equation}
where $U_{n}(x)$ is the Chebyshev polynomials of the second kind (See Appendix A in Ref.~\cite{rn14}). Substituting $\alpha_{1,R}$, $\alpha_{N,R}$ into Eq.~(\ref{a8}), we arrive at $2x\alpha_{1,L}=\alpha_{2,L}+\alpha_{N,L}$. Analogously, substituting $\alpha_{N,R}$, $\alpha_{N-1,R}$ into $a\alpha_{N,L} +b\alpha_{N,R}=u\alpha_{N-1,R}$ leads to $2x\alpha_{N,L}=\alpha_{N-1,L}+\alpha_{1,L}$. Utilizing Eq.~(\ref{a9}) to eliminate $\alpha_{N,L}$ and $\alpha_{N-1,L}$ , we obtain two independent equations for $\alpha_{2,L}$ and $\alpha_{1,L}$,
\begin{equation}\label{a10}
 [1+U_{N-2}(x)]\alpha_{2,L} =  [2x+U_{N-3}(x)]\alpha_{1,L}
\end{equation}
and
\begin{equation}\label{a11}
\begin{aligned}
( 2xU_{N-2}(x)-U_{N-3}(x) ) \alpha_{2,L} & \equiv U_{N-1}(x)\alpha_{2,L}  = [2xU_{N-3}(x)-U_{N-4}(x)+1]\alpha_{1,L} \\
&\equiv [U_{N-2}(x)+1]\alpha_{1,L}. \ \text{(A7) in Ref.~\cite{rn14} is used}.
\end{aligned}
\end{equation}

Eqs.~(\ref{a10}) and (\ref{a11}) should have nonzero solutions, the determinant of the four coefficients equals to 0, which leads to,
\begin{equation}\label{a12}
[U_{N-2}(x)+1]^2-[2x+U_{N-3}(x)]U_{N-1}(x)=0.
\end{equation}
According to the Chebyshev identities (A8) and (A10) in Ref.~\cite{rn14}, we have $xU_{N-1}(x)-U_{N-2}(x)=T_{N}(x)$ and $U^2_{N-2}(x)-U_{N-1}(x)U_{N-3}(x)=U_0(x)=1$. Thus Eq.~(\ref{a12}) can be simplified as,
\begin{equation}\label{a13}
T_N(x)=1.
\end{equation}
Here, $T_n(x)$ is the Chebyshev polynomial of the first kind (see Appendix A in Ref.~\cite{rn14}). The $N$ solutions of the above equation can be represented as the following simple trigonometric function,
\begin{equation}\label{a14}
x_j=\cos\theta_j, \theta_j = \frac{2j\pi}{N}.\  \forall \ N, \ j=1,2,...,N.
\end{equation}
Using the mapping relation $u=bx\pm i\sqrt{1-b^2x^2}$, the $2N$ eigenvalues of $\hat{U}$ are given by,
\begin{equation}\label{a15} \small
u_{j,\pm}=bx_j\pm i\sqrt{1-b^2x_j^2}, x_j=\cos\theta_j, \theta_j= \frac{2j\pi}{N}, j\in [1,N].
\end{equation}

\subsection{Eigenstates}
Now we analyze the eigenstates $|\Psi\rangle$. According to Eqs.~(\ref{a6}) and (\ref{a9}), the right components $\alpha_{j,R}^s$ can be written as a function of $\alpha_{2,L}$ and $\alpha_{1,L}$,
\begin{equation}\label{a16}
\begin{aligned}
\alpha_{j,R} = &\frac{1}{a}\{ [bU_{j-2}(x)-uU_{j-1}(x)]\alpha_{2,L} - [bU_{j-3}(x)-uU_{j-2}(x)]\alpha_{1,L}   \}
\end{aligned}
\end{equation}
When $T_N(x)=1$, the four coefficients of $\alpha_{2,L}$ and $\alpha_{1,L}$ in Eqs.~(\ref{a10}) and (\ref{a11}) are zero. Here, for the discrete eigenvalues we set $\alpha_{2,L}(x_j)=B(a,b,\theta_j)e^{2i\theta_j}$ and $\alpha_{1,L}(x_j)=B(a,b,\theta_j)e^{i\theta_j}$. According to Bloch theorem and Eq.~(\ref{a9}), it is easy to find a general Bloch ansatz solution for $\alpha_{k,L}^s(x_j)$,
\begin{equation} \label{a17}
\alpha_{k,L}(x_j)=B(a,b,\theta_j)e^{ik\theta_j}.
\end{equation}
Likewise, according to Eq.~(\ref{a16}), we find a general solution for $\alpha_{k,R}(x_j)$,
\begin{equation}\label{a18}
\alpha_{k,R}(\theta_j)|_{u_{j,\pm}}= \frac{\mp i}{a} e^{i\theta_j}\Big(\sqrt{1-b^2\cos^2\theta_j} \pm b\sin\theta_j  \Big) B(a,b,\theta_j)e^{ik\theta_j}.
\end{equation}
Here, the factor $B(a,b,\theta_j)$ can be determined by the normalization condition $\sum_{k,J}|\alpha_{k,J}^m(x_j)|^2=1$. After some algebraic calculus, we obtain,
\begin{equation}\label{a19}
|B(a,b,\theta_j)|_{u_{j,\pm}}= \frac{a}{\sqrt{N[a^2+(\sqrt{1-b^2\cos^2\theta_j}\pm b\sin\theta_j)^2}]}.
\end{equation}
Thus, we have obtained all the orthonormalized eigenstates for $\hat{U}$.

\end{document}